\documentclass[letterpaper,twocolumn,10pt]{article}
\usepackage{usenix}

\usepackage{tikz}
\usepackage{amsmath}

\usepackage{tcolorbox}
\usepackage{listings}
\usepackage{xcolor}
\usepackage{algorithm}
\usepackage{algpseudocode}
\usepackage{graphicx}

\usepackage{enumitem}

\usepackage{tikz}

\usepackage{mdframed}
\usepackage{xcolor}
\usepackage{pifont}
\newcommand{\cmark}{\ding{51}}
\newcommand{\xmark}{\ding{55}}
\definecolor{verylightgray}{gray}{0.97}

\usepackage{multirow}
\usepackage[flushleft]{threeparttable}
\usepackage{tabularx} % Load the tabularx package

\usepackage{tikz}
\usepackage{amsmath}

\usepackage{soul}

\usepackage{microtype}
\usepackage{color, colortbl}
\usepackage{hyperref}
\usepackage{float}
\floatstyle{plaintop}
\restylefloat{table}
\usepackage{xurl}
\usepackage{caption} 
\usepackage{comment}

\newcommand{\shortsectionBf}[1]{\vspace{2.5pt}
\noindent {\bf #1}
}

\usepackage{xspace}
\newcommand{\sys}{{\textsc{\small{MARS}}}\xspace}

\usepackage{microtype}
\usepackage{color, colortbl}
\usepackage{hyperref}
\usepackage[scaled=.7]{beramono}
\usepackage{float}
\floatstyle{plaintop}
\restylefloat{table}
\usepackage{xurl}
\usepackage{caption} 

\usepackage{filecontents}

\begin{document}
%-------------------------------------------------------------------------------

%don't want date printed
\date{}

% make title bold and 14 pt font (Latex default is non-bold, 16 pt)
\title{MARS: Detecting Unauthorized Variable Manipulations in \\ Multi-Application PLC Runtimes}

\author{
{\rm Syed Ghazanfar Abbas, Dongyan Xu}\\
{\it Purdue University,}\\
{\it \{abbas4, dxu\}@purdue.edu}
}

\maketitle

\begin{abstract}
Programmable Logic Controllers (PLCs) increasingly run multiple applications alongside the main control program, with shared access to PLC variables. Yet, Industrial Control System (ICS) defenses primarily detect malicious updates by checking whether variable values violate expected bounds, without considering which application performed the update. A malicious application can exploit this gap by modifying variables within normal bounds while still driving the physical process toward an unsafe state. Even when such manipulation is detected, operators cannot identify the responsible application because PLCs do not associate variable updates with application identity. 

We present \sys, an automated framework for application-level authorization and attribution of PLC variable manipulations. \sys profiles applications on an isolated virtual PLC (vPLC) to derive application-specific variable-access policies and uses a shadow vPLC during operation to attribute production-PLC updates to individual applications without instrumenting the production controller. \sys also detects manipulations that occur only on the production PLC and therefore have no corresponding update on the shadow vPLC. We evaluate \sys on manufacturing, chemical, and water-treatment systems against attacks in which unauthorized applications manipulate PLC variables while remaining within normal bounds. Our results show that \sys detects these manipulations and identifies the responsible application.

%We present \sys, an automated framework for application-level authorization and attribution of PLC variable manipulations. \sys profiles applications on an isolated virtual PLC (vPLC) to derive application-specific variable-access policies and uses a shadow vPLC during operation to attribute production-PLC updates to individual applications without instrumenting the production controller. \sys also detects manipulations that occur only on the production PLC and therefore have no corresponding update on the shadow vPLC. We evaluate \sys on manufacturing, chemical, and water-treatment systems against attacks in which unauthorized applications manipulate PLC variables while remaining within normal bounds. Our results show that \sys detects these manipulations and identifies the responsible application.

%variable-access policies → variable writes

\end{abstract}

\section{Introduction}
\label{sec:introduction}

Programmable Logic Controllers (PLCs) have been repeatedly targeted in attacks against critical infrastructure, where adversaries manipulated PLC variables to influence physical processes. In 2026, the National Security Agency (NSA) and partner agencies warned that cyber actors were actively targeting PLCs across critical-manufacturing, energy, water, and chemical sectors using AI-generated exploitation scripts disguised as legitimate monitoring tools~\cite{nsa2026plc}. Earlier incidents demonstrated the physical consequences of PLC-variable manipulation: Stuxnet altered centrifuge-control variables on Siemens PLCs while concealing the resulting physical behavior from plant monitoring~\cite{falliere2011w32}, and the Oldsmar water-treatment intrusion attempted to modify a chemical-dosing setting to a dangerous level~\cite{florida}. Similar attacks have affected power distribution~\cite{attack41} and manufacturing systems~\cite{maggi2020attacks, williams2022taxonomy}.

\shortsectionBf{Multi-Application PLCs.}
Modern PLCs are increasingly becoming multi-application platforms. Alongside the main control program~\cite{tiegelkamp2010iec}, operators deploy vendor- and third-party applications within the PLC runtime for visualization, diagnostics, logging, maintenance, communication, and cloud connectivity~\cite{@storePLCNext,@storeSiemens,@storeCodesys}. These applications extend PLC functionality, but they can also access variables used by the control program to make control decisions~\cite{@SAIN}. If one of these applications is vulnerable or compromised, an attacker can modify such variables without altering the control program itself~\cite{pickren2024compromising}. The unchanged control program can then consume the manipulated value and drive the physical process accordingly.

\shortsectionBf{Application-Agnostic ICS Defenses.}
Existing ICS defenses largely detect PLC-variable manipulation by determining whether observed values or process behavior violate expected constraints~\cite{cheng2017orpheus,feng2019systematic,@SAIN,hadvziosmanovic2014through,vetplc}. Such defenses detect abnormal process behavior, but generally reason about the resulting value or state rather than the runtime application that produced the modification. In a multi-application PLC, the same variable may be accessible to multiple co-running applications, while each application has a different authorization scope. Observing a valid variable value therefore does not establish that the application responsible for the write was permitted to modify that variable.

This gap affects both attack detection and operator response. A compromised application may change a valid setpoint, mode, or control variable to alter production, dosing, or actuator behavior while keeping the resulting value within expected ranges. Even when such a manipulation eventually produces an observable anomaly and is detected, the operator may still be unable to determine which co-running application caused the modification because commodity PLC runtimes generally do not expose writer identity. Identifying the responsible runtime application is therefore necessary not only for detecting unauthorized writes that remain value-valid, but also for diagnosing the source of detected manipulations and supporting targeted response.

\shortsectionBf{Approach.}
We present \sys\footnote{The acronym \textsc{MARS} stands for ``Multi-App Runtime Security''.}, an automated framework for detecting unauthorized PLC-variable writes and recovering their runtime-application origin in multi-application PLC runtimes. \sys uses instrumented virtual PLCs (vPLCs) to recover application-labeled variable writes and runtime-application execution information without modifying the live production controller.

Before deployment, \sys runs the PLC project and its runtime applications on an isolated instrumented vPLC. This allows \sys to observe which applications write monitored variables and validate those writes against an operator-supplied policy specifying which applications are authorized to modify each variable.

During plant operation, \sys runs an instrumented \emph{shadow vPLC} alongside the live PLC. The live PLC provides the variable modifications that actually occur, while the shadow provides runtime-application execution information that the live PLC does not expose. \sys aligns the two controllers at control-cycle granularity rather than assuming timestamp-level lockstep.

The central challenge is that a malicious write may occur only on the live PLC and therefore have no corresponding write on the shadow. For each live-PLC modification, \sys first checks for a corresponding application-labeled write on the shadow vPLC. If none exists, \sys uses nearby shadow-vPLC application executions as bounded candidate evidence of the possible writer. It reports a unique application only when the available evidence supports one; otherwise, it preserves ambiguity and reports only the authorization conclusion supported by the candidate set.

We evaluate \sys on manufacturing, chemical-processing, and water-treatment systems using 15 application-level attacks in which compromised runtime applications modify PLC variables while keeping the written values within their legitimate domains. \sys detects all 15/15 attacks during pre-deployment validation. In the harder setting, the malicious branch executes only on the live PLC and does not appear on the shadow. Across 150 repeated live-PLC executions, \sys identifies all evaluated authorization violations and uniquely attributes the responsible runtime application in 140/150 runs (93.3\%); the remaining 10 executions are reported with ambiguous origin, with no incorrect unique attribution. Across 59,810 held-out benign PLC-variable updates, \sys produces no false unauthorized-write alerts. Median application-aware decision latency ranges from 24.3--32.7\,ms across the three evaluated platforms.

In summary, this paper makes the following contributions:

\begin{itemize}[leftmargin=*,topsep=2pt,itemsep=1pt]

\item \textit{Application-level PLC-variable authorization.}
We identify a security gap in multi-application PLC runtimes where a PLC-variable value can remain legitimate while the runtime application performing the write is not authorized to modify that variable.

\item \textit{Black-box pre-deployment write validation.}
We recover application-labeled PLC-variable writes from closed-source runtime applications on an instrumented vPLC and validate them against independently supplied application write policies, enabling unauthorized writes to be detected before deployment.

\item \textit{Evidence-based attribution of live-PLC-only writes.}
We design a shadow-vPLC technique that detects PLC-variable modifications that occur only on the live PLC and uses bounded runtime-application execution evidence to attribute their origin when the evidence is sufficient, while explicitly preserving ambiguity otherwise.

\item \textit{Cross-platform evaluation.}
We evaluate \sys on three ICS testbeds using 15 application-level attacks. \sys detects all evaluated attacks, uniquely attributes the responsible application in 140 of 150 live-PLC executions (93.3\%), reports the remaining 10 as ambiguous with no incorrect unique attribution, produces no false unauthorized-write alerts across 59,810 benign updates, and achieves 24.3--32.7\,ms median application-aware decision latency.

\end{itemize}

\section{Threat Model}
\label{sec:threatmodel}

\begin{figure}[t]
\centering
\includegraphics[width=.80\columnwidth]{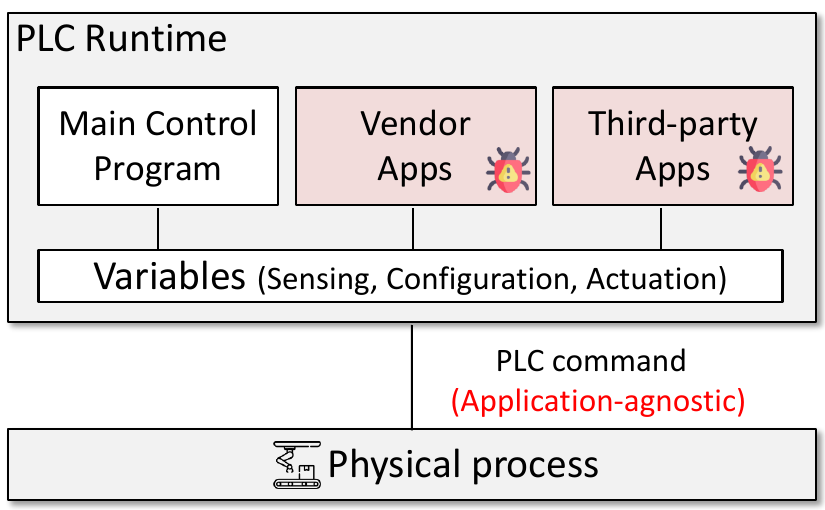}
\caption{Multi-app PLC runtime and MARS threat model.}
\label{fig:threatmodel}
\end{figure}

\shortsectionBf{System.}
We consider a multi-application PLC runtime in which the control program executes together with vendor- and third-party applications for functions such as diagnostics, visualization, logging, maintenance, and communication~\cite{@storePLCNext,@storeSiemens,@storeCodesys}. These applications may access PLC variables that are also used by the control program (Figure~\ref{fig:threatmodel}).

\shortsectionBf{Operator's perspective.}
\sys is used by the PLC operator or security engineer. The operator has the authorized PLC project and application packages, can execute the same software on isolated vPLCs, and supplies the write-access policy for monitored variables. The policy can be derived from engineering configuration, site security policy, application documentation, and operator knowledge; the final authorization remains under operator control. A developer-provided manifest or an observed profiling trace cannot silently expand an application's authority.

\shortsectionBf{Attacker capabilities.}
We consider a compromised, vulnerable, or malicious runtime application~\cite{hou2019understanding,pickren2024compromising}. The attacker can execute application code and read or write PLC variables reachable through the runtime. The attacker can choose in-range values and time a write relative to other runtime applications. The attacker does not need to modify the main control program. In the live-PLC-only setting, the compromised application can continue its normal runtime execution on the isolated/shadow vPLC while its malicious branch is triggered only on the plant PLC.

\shortsectionBf{Trusted components and scope.}
We trust the main control program, the operator-supplied write-access policy, the vPLC/shadow environment, and the external MARS host. We do not consider physical field-device attacks, compromise of these trusted components, or malicious use of a variable that an application is already authorized to write. The plant PLC and shadow vPLC are assumed to contain the same deployed application set. If unknown software exists only on the plant PLC, MARS can observe unexplained behavior but cannot assign it a known shadow identity.

\section{Motivation}
\label{sec:motivation}

We use the manufacturing testbed to demonstrate an application-level authorization violation in which an unauthorized runtime application writes a legitimate value to a PLC variable, causing an unintended physical effect on the plant.

\begin{figure}[t]
\centering
\includegraphics[width=.75\columnwidth]{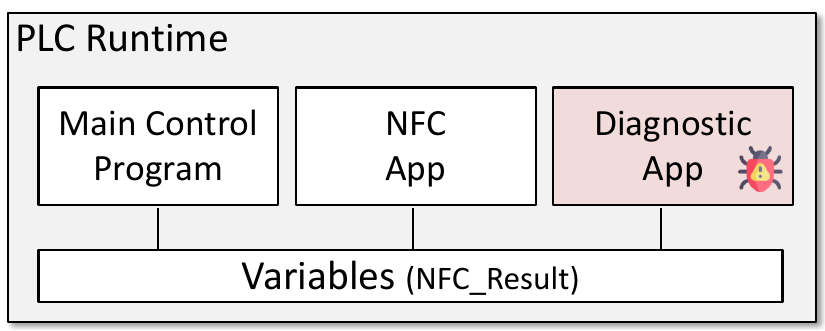}
\caption{Multiple applications in a manufacturing PLC runtime sharing access to
\texttt{NFC\_Result}.}
\label{fig:nfc_motivation}
\end{figure}

\shortsectionBf{Manufacturing Testbed.}
Our manufacturing testbed implements an automated material-handling process in which a robotic arm transports workpieces between processing and storage stations~\cite{@fp}. Before a workpiece is routed, it is classified using NFC information. As shown in Figure~\ref{fig:nfc_motivation}, the PLC runtime contains the main control program together with an NFC application and a diagnostic application, all of which access the shared PLC variable \texttt{NFC\_Result}.

The control program uses \texttt{NFC\_Result} to determine how the workpiece should be routed. During classification, \texttt{NFC\_Result} can take the values \texttt{WAIT}, \texttt{VALID}, or \texttt{DISCARD}. The NFC application is authorized to update this variable with the classification result, whereas the diagnostic application reads the same variable only for monitoring and is not authorized to modify it.

\shortsectionBf{Unauthorized In-Range Write.}
We compromise the diagnostic application while classification of a valid workpiece is still in progress. At this point, \texttt{NFC\_Result=WAIT}, and the control program is waiting for the NFC application to produce the final classification. Before that result is generated, the compromised diagnostic application writes \texttt{NFC\_Result=DISCARD}.

\texttt{DISCARD} is a legitimate classification value that the control program is designed to accept. The control program therefore follows its unchanged logic and routes the workpiece to the discard station. The NFC application later completes the classification and produces \texttt{VALID}, but the routing decision has already been executed. The attack succeeds without modifying the control program or writing an abnormal value to \texttt{NFC\_Result}.

\shortsectionBf{Why Value-Level Monitoring Is Insufficient.}
The key distinction is which application performs the write. In a legitimate execution, \texttt{NFC\_Result=DISCARD} may be produced by the NFC application after classification. In the attack, the diagnostic application produces the same value before classification completes. From the PLC-variable value alone, the two writes are indistinguishable: both contain a valid \texttt{DISCARD} value, but only the NFC application is authorized to write \texttt{NFC\_Result} in this setting.

The physical consequence depends on when the unauthorized write occurs. Here, the diagnostic application writes a terminal classification while the control program is awaiting the NFC result, causing the manipulated value to be consumed before the legitimate classification arrives. Regardless of whether a later write changes the physical outcome, the unauthorized modification still violates the application's write policy.

This example exposes the information missing from conventional PLC-variable monitoring. Observing the transition \texttt{WAIT}$\rightarrow$\texttt{DISCARD} reveals the variable change, but not whether the NFC application or the diagnostic application was responsible for it. Detecting this class of manipulation therefore requires application-level evidence in addition to the observed PLC-variable value.

\begin{figure*}[t!]
\center
\includegraphics[width=1\textwidth]{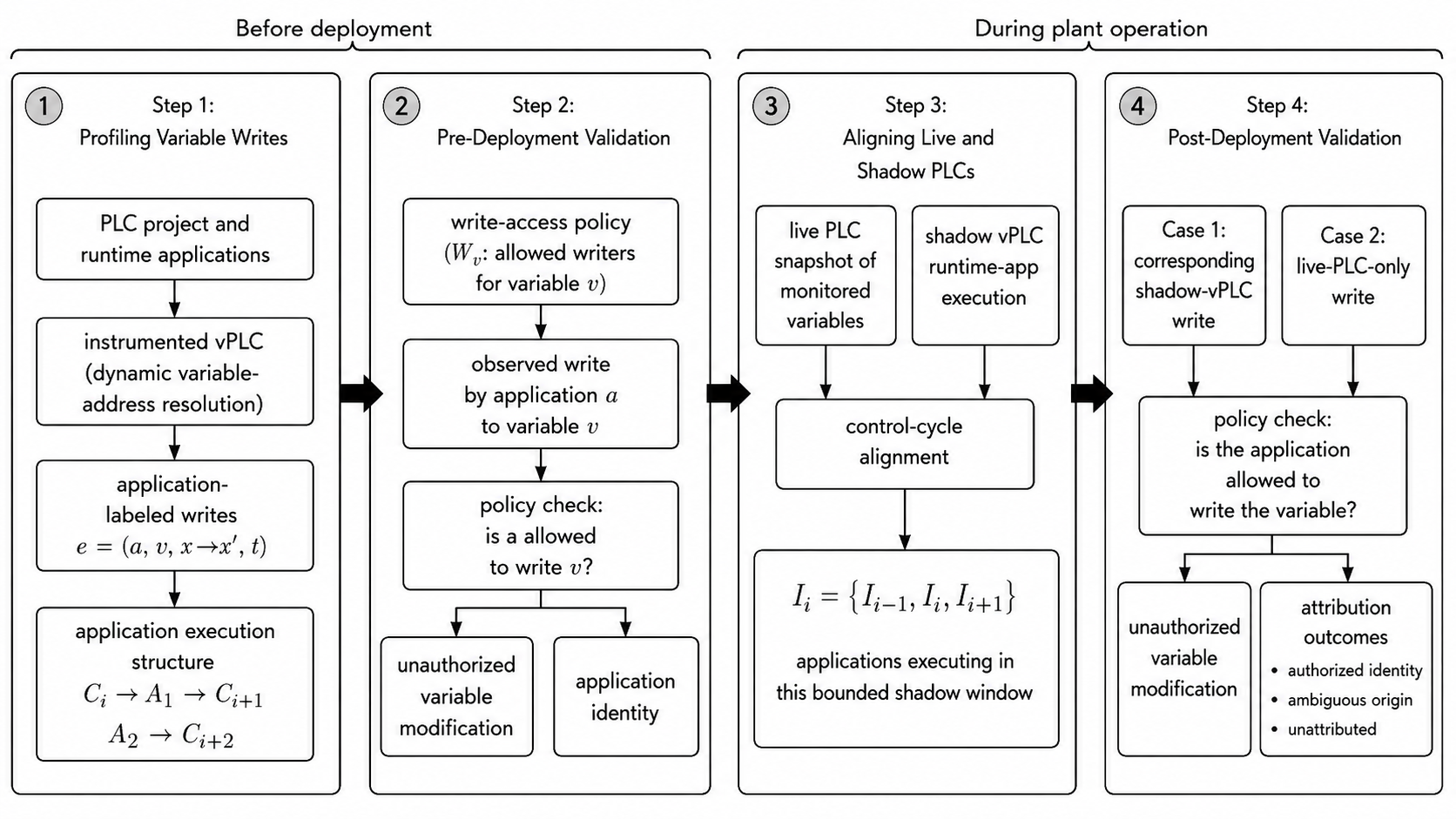}
\caption{Overview of \sys architecture.}
\label{fig:System-Architecture}
\end{figure*}

\section{\sys Overview}
\label{sec:overview}
\sys determines whether a PLC-variable modification is authorized and, when
the available evidence permits, identifies the runtime application associated
with the modification. As shown in Figure~\ref{fig:System-Architecture}, \sys operates in
two phases: pre-deployment validation and post-deployment validation. Before deployment, \sys runs the PLC project and its runtime
applications on an isolated instrumented virtual PLC (vPLC). This environment
allows \sys to recover application-labeled PLC-variable writes and to record
how runtime applications execute relative to the periodically executing main
control program. During plant operation, \sys runs an instrumented shadow vPLC
alongside the live PLC. The live PLC remains unchanged and provides a snapshot
of the monitored PLC variables after each control cycle, while the shadow vPLC
provides application-level information that the live PLC does not expose.

\shortsectionBf{Inputs.}
\sys takes as input the PLC project and runtime applications, the monitored PLC
variables, and operator-supplied application write-access constraints. For each
monitored variable $v$, the policy defines
\[
W_v=\{a\mid a\text{ is permitted to modify }v\}.
\]

\shortsectionBf{Design challenges.}
\sys addresses three challenges.

\emph{C1: Recovering application-level evidence.}
The live PLC exposes monitored PLC-variable changes but does not identify the
runtime application that performed a write. \sys therefore uses an instrumented
vPLC to recover application-labeled variable writes and runtime-application
execution information.

\emph{C2: Separating behavior from authorization.}
Observed application behavior alone cannot determine whether a write is
authorized. \sys therefore validates application writes against the
operator-supplied write-access constraints $W_v$.

\emph{C3: Attributing live-PLC-only writes.}
A malicious application may perform an unauthorized write only on the live PLC,
so the same write may not appear on the shadow vPLC. Because the live PLC
provides one monitored-variable snapshot after each control cycle, \sys
localizes the visible change to one control-cycle interval. \sys first examines
the corresponding shadow-vPLC control cycle and, when needed, the immediately
preceding and following cycles. The applications observed in this bounded
window form the candidate set for attribution. \sys reports a unique application only
when the available evidence identifies a single candidate; otherwise, it
reports the remaining candidate set.

\shortsectionBf{Workflow.}
\sys consists of four steps. Step~1 profiles application-labeled PLC-variable
writes and runtime-application execution structure on the instrumented vPLC.
Step~2 performs pre-deployment validation by checking each observed write
against the operator-supplied write policy. Step~3 aligns live and shadow PLC
control cycles. Step~4 performs post-deployment validation for both live transitions with
corresponding shadow-vPLC writes and transitions observed only on the live PLC.

\section{\sys Design}
\label{sec:design}

\subsection{Step 1: Profiling Variable Writes}
\label{lab:step1}

Step~1 executes the PLC project and runtime applications on an isolated
instrumented vPLC. \sys collects two forms of information used by the later
validation steps: application-labeled PLC-variable writes and runtime-application executions relative to the main control program.

\shortsectionBf{Dynamic variable-address resolution.}
Runtime addresses of PLC variables may change after project download, online
change, or runtime restart. \sys therefore resolves monitored variables
dynamically instead of relying on fixed offsets. A read-only Structured Text
resolver executes on the instrumented vPLC and uses the IEC~61131-3
\texttt{ADR} operator to obtain the current runtime address of each monitored
variable~\cite{tiegelkamp2010iec,Siemens2018scl}. The resolver obtains addresses only and does not modify monitored
values.

\begin{figure}[t]
\centering
\includegraphics[width=.96\columnwidth]{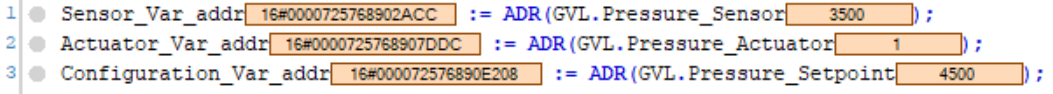}
\caption{Resolving PLC-variable runtime addresses on the instrumented vPLC.}
\label{fig:addr}
\end{figure}

\shortsectionBf{Application-labeled writes.}
Using the resolved address map, an external runtime monitor observes writes to
the monitored PLC variables. For each write, \sys records the application,
target variable, value transition, and observation time:
\[
e=\langle a,v,x\!\rightarrow\!x',t\rangle.
\]
This application-labeled write trace is used directly in Step~2 for
pre-deployment validation.

\sys does not treat profiling as a complete list of variables that an
application can ever modify. A compromised application may remain benign
during profiling and perform a new unauthorized write only after deployment.
The absence of a write during profiling is therefore never used to exclude an
application from post-deployment attribution.

\shortsectionBf{Application execution structure.}
The main control program executes periodically as the recurring control task.
Other runtime applications execute independently: some are periodic, while others are triggered by runtime or external
events~\cite{canedo2014high,diedrich2004function}. Consequently, an
application may execute between some main-control-program executions but not
others.

The instrumented vPLC records these application executions relative to the
recurring executions of the main control program. For example, an observed
segment may be
\[
C_i \rightarrow A_1 \rightarrow C_{i+1}
    \rightarrow A_2 \rightarrow C_{i+2},
\]
where $C_i$, $C_{i+1}$, and $C_{i+2}$ denote consecutive executions of the
main control program, and $A_1$ and $A_2$ denote other runtime applications.
This information is later used to determine which applications executed during
a particular shadow-vPLC control cycle.

\subsection{Step 2: Pre-Deployment Validation}
\label{lab:step2}

Before deployment, the instrumented vPLC directly identifies the application
that performs each monitored-variable write. \sys therefore detects and
attributes unauthorized variable modifications by checking each observed write
against the operator-supplied policy.

For an observed write by application $a$ to variable $v$,
\[
a\in W_v
\]
means that the write is authorized, while
\[
a\notin W_v
\]
means that the write is unauthorized. Because the instrumented vPLC directly provides $a$, \sys reports an
unauthorized variable modification together with the application that performed
the write.

For example, suppose only the NFC application is permitted to modify
\texttt{NFC\_Result}:
\[
W_{\texttt{NFC\_Result}}=\{\text{NFC application}\}.
\]
If the Diagnostic application writes
\texttt{NFC\_Result}: \texttt{WAIT}$\rightarrow$\texttt{DISCARD}, \sys reports
an unauthorized modification of \texttt{NFC\_Result} and attributes it to the
Diagnostic application, even though \texttt{DISCARD} is a legitimate value of
the variable.

Observing an application write during profiling never expands $W_v$. Thus, an
unauthorized write exercised before deployment is reported as a violation
rather than learned as legitimate behavior.

\subsection{Step 3: Aligning Live and Shadow PLCs}
\label{lab:step3}

During plant operation, the live PLC and shadow vPLC are configured with the
same PLC project, runtime applications, control-task configuration, and
relevant sensor and supervisory inputs. The same deployed runtime applications execute on both controllers. In the
live-PLC-only setting, the compromised application continues its normal
runtime execution on the shadow vPLC while its malicious branch executes only
on the live PLC. \sys aligns the two controllers at the
main-control-program cycle level rather than by individual application
timestamps.

\shortsectionBf{Live-PLC control-cycle snapshots.}
The live PLC provides a snapshot of the monitored variables after each main
control cycle. Let $L_i$ denote the snapshot collected after control cycle
$C_i$. For two consecutive snapshots $L_i$ and $L_{i+1}$, \sys determines
which visible values changed:
\[
\Delta L_i=\{v\mid L_i(v)\neq L_{i+1}(v)\}.
\]
If $v$ changes from $x$ in $L_i$ to $x'$ in $L_{i+1}$, the visible transition
occurred during the control-cycle interval
\[
I_i=(C_i,C_{i+1}].
\]
The live PLC does not reveal which runtime application performed the write or
the order of multiple writes that occurred within $I_i$.

\shortsectionBf{Control-cycle alignment.}
Because the live PLC and shadow vPLC use the same control-task configuration
and receive the same relevant external inputs, \sys associates live control
interval $I_i$ with the corresponding main-control-program interval on the
shadow vPLC. The two controllers need not execute individual runtime
applications at identical wall-clock times; the recurring main-control-program
cycles provide the common reference.

For a live transition localized to $I_i$, \sys first examines the corresponding
shadow interval $I_i$. Because periodic or event-driven runtime applications
need not execute in the same relative control cycle on the two PLCs, \sys
allows a bounded displacement around the corresponding shadow interval. In our
design, we use the immediately preceding and following cycles:
\[
\mathcal{I}_i=\{I_{i-1},I_i,I_{i+1}\}.
\]
We evaluate the adequacy of this window experimentally in
Section~\ref{sec:evaluation}. This center-first search prioritizes the
corresponding control cycle while tolerating small execution displacement
between the two PLCs.

For live-PLC-only attribution, \sys records the applications that execute
within this bounded shadow window:
\[
E_i=
\bigcup_{j=i-1}^{i+1}
\{a\mid a\text{ executes during shadow control interval }I_j\}.
\]
$E_i$ is used in Step~4 only when no matching application-labeled write is
found after the center-first search.

\subsection{Step 4: Post-Deployment Validation}
\label{lab:step4}

Post-deployment validation examines variable transitions between consecutive
live-PLC snapshots. For each live transition
$u=(v,x\!\rightarrow\!x')$ occurring in control interval $I_i$, \sys first
searches the corresponding shadow interval for the same transition
(Line~\ref{algline:center}). If no match exists, it searches the immediately
preceding and following intervals (Line~\ref{algline:neighbors}). This
center-first search produces two cases.

\begin{algorithm}[t!]
\caption{\sys post-deployment validation}
\label{alg:mars_post}
\begin{algorithmic}[1]
\Require Live snapshots $L$, shadow trace $S$, policy $W$

\For{$i=1$ to $|L|-1$}
    \State $D \gets \textsc{Transitions}(L_i,L_{i+1})$
    \State $\mathcal{I}_i \gets \{I_{i-1},I_i,I_{i+1}\}$

    \ForAll{$u=(v,x\!\rightarrow\!x')\in D$}
        \State $M \gets \textsc{ShadowMatches}(u,I_i,S)$
        \label{algline:center}

        \If{$M=\emptyset$}
            \State $M \gets \textsc{ShadowMatches}(u,\mathcal{I}_i\setminus\{I_i\},S)$
            \label{algline:neighbors}
        \EndIf

        \If{$|M|=1$}
            \State $a \gets \textsc{Application}(M)$
            \State \textsc{Report}$(u,a,\textsc{Policy}(a,W_v))$
            \label{algline:matched}

        \ElsIf{$|M|>1$}
            \State \textsc{Report}$(u,M,\textsc{MultipleMatches})$

        \Else
            \State $E_i \gets \textsc{AppsInWindow}(\mathcal{I}_i,S)$
            \label{algline:candidates}

            \If{$E_i=\emptyset$}
                \State \textsc{Report}$(u,\textsc{Unattributed})$

            \ElsIf{$|E_i|=1$}
                \State $a \gets \textsc{Only}(E_i)$
                \State \textsc{Report}$(u,a,\textsc{Policy}(a,W_v))$

            \ElsIf{$E_i\cap W_v=\emptyset$}
                \State \textsc{Report}$(u,E_i,\textsc{Unauthorized})$
                \label{algline:unauth-amb}

            \Else
                \State \textsc{Report}$(u,E_i,\textsc{Ambiguous})$
                \label{algline:amb-origin}
            \EndIf
        \EndIf
    \EndFor
\EndFor
\end{algorithmic}
\end{algorithm}

\shortsectionBf{Case 1: Corresponding shadow-vPLC write.}
Suppose variable $v$ changes from $x$ to $x'$ between $L_i$ and $L_{i+1}$,
and \sys finds the same transition through the center-first shadow-vPLC search.
Because the shadow vPLC is instrumented, each matching shadow write carries
the identity of the application that produced it.

If exactly one shadow write matches the live transition, \sys obtains the
application identity $a$ associated with that shadow write and checks whether
\[
a\in W_v.
\]
If $a\in W_v$, the modification is authorized; otherwise, \sys reports an
unauthorized modification attributed to $a$. If multiple shadow writes match
the same live transition, \sys retains the matching applications rather than
forcing a unique attribution.

For example, if both the live PLC and shadow vPLC contain
\texttt{NFC\_Result}: \texttt{WAIT}$\rightarrow$\texttt{VALID}, and the
matching shadow write is labeled as originating from the NFC application,
\sys validates the NFC application against
$W_{\texttt{NFC\_Result}}$. If the matching write instead originates from
Diagnostic and Diagnostic is not permitted to modify the variable, \sys
reports an unauthorized modification attributed to Diagnostic.

\shortsectionBf{Case 2: Live-PLC-only write.}
If no matching shadow write exists within
\[
\mathcal{I}_i=\{I_{i-1},I_i,I_{i+1}\},
\]
\sys treats the transition as live-PLC-only. Because the live PLC does not
expose the writer identity, \sys obtains the applications that executed in the
corresponding shadow window:
\[
E_i=
\bigcup_{j=i-1}^{i+1}
\{a\mid a\text{ executes during shadow control interval }I_j\}.
\]
$E_i$ represents candidate applications whose shadow execution is temporally
consistent with the observed live-only transition. Candidate membership does
not require an application to have previously written the target variable
during profiling.

If $E_i$ contains exactly one application $a$, \sys uniquely attributes the
transition to $a$ and evaluates $a$ against $W_v$. For example, if
\[
W_{\texttt{NFC\_Result}}=\{\text{NFC application}\}
\]
and
\[
E_i=\{\text{Diagnostic}\},
\]
then \sys reports an unauthorized live-PLC-only modification of
\texttt{NFC\_Result} and attributes it to Diagnostic.

If multiple applications remain, \sys does not force a unique identity. When
\[
E_i\cap W_v=\emptyset,
\]
all candidate applications are unauthorized for $v$, so \sys reports an
unauthorized modification with ambiguous application attribution
(Line~\ref{algline:unauth-amb}). Otherwise, at least one authorized and one
unauthorized candidate may remain, so \sys reports the complete candidate set
and reports the origin as ambiguous (Line~\ref{algline:amb-origin}). If
$E_i=\emptyset$, \sys reports the live-PLC-only transition as unattributed.

\shortsectionBf{Attribution outcomes.}
For a live-PLC-only transition, \sys reports a unique application only when
$|E_i|=1$. If $|E_i|>1$, it reports the complete candidate set and only the
authorization conclusion jointly supported by the candidate evidence and
$W_v$. If $E_i=\emptyset$, it reports the unexplained live-PLC-only transition
without assigning a writer identity.

\section{Evaluation}
\label{sec:evaluation}

\subsection{Experimental Setup}
\label{subsec:setup}

We evaluate \sys on three industrial testbeds: a Fischertechnik manufacturing
plant (MP)~\cite{@fp}, a chemical-processing system (CP)~\cite{@cp}, and a water-treatment system
(WP)~\cite{@wp}. MP is a modular production line with conveyors, robotic arms, NFC-based
workpiece handling, and diverse sensors and actuators. CP includes pumps,
valves, mixers, and pressure/flow control loops. WP includes chemical dosing,
pump sequencing, and tank-level regulation. Table~\ref{tab:testbeds} summarizes
the physical and virtual controllers used in the evaluation. These testbeds exercise three distinct evaluated PLC/vPLC stacks---PLCnext/
VPLCNEXT, Unipi/OpenPLC, and CODESYS Control---while the same \sys attribution
and policy-validation logic is used across them; platform-specific
instrumentation is confined to the virtual PLC side.

\begin{table}[t!]
\centering
\footnotesize
\caption{Industrial testbeds used in evaluation.}
\label{tab:testbeds}

\renewcommand{\arraystretch}{1.12}
\setlength{\tabcolsep}{3.5pt}

\begin{threeparttable}
\begin{tabular}{|c|p{2.75cm}|p{2.65cm}|}
\hline
\textbf{Testbed} & \textbf{Physical PLC} & \textbf{Virtual PLC} \\
\hline
MP & PLCnext Control AXC F 2152~\cite{PLCNextFirmwareupdat} & VPLCNEXT Control 2000~\cite{VPLCNext} \\
\hline
CP & Unipi Neuron L203~\cite{@unipi} & OpenPLC~\cite{alsabbagh2024investigating} \\
\hline
WP & Raspberry Pi 4 running CODESYS Control for Linux SL~\cite{@codesysPLC} &
CODESYS Control SL~\cite{@codesysVPLC} \\
\hline
\end{tabular}

\begin{tablenotes}[flushleft]
\scriptsize
\item MP: manufacturing plant; CP: chemical plant; WP: water-treatment plant.
\end{tablenotes}

\end{threeparttable}
\end{table}

\shortsectionBf{Trace collection.}
We execute the PLC project and its runtime applications on an isolated
instrumented vPLC to recover application-labeled PLC-variable writes and the
mapping needed to identify runtime-application execution on the shadow vPLC.
This profiling supports pre-deployment validation and recovery of
application-level execution evidence. Post-deployment attribution instead uses
current shadow-vPLC runtime-application execution within a bounded
control-cycle window, rather than a stored history of prior writes. Application write-access constraints
$W_v$ are supplied independently of profiling; an unauthorized write observed
during profiling is therefore reported rather than learned as legitimate
authorization.

\shortsectionBf{Attack suite.}
Attack generation is used only to construct the evaluation workload and is not
part of \sys. We construct 15 attacks, five per testbed, from benign
PLC-variable traces and manually validate each candidate on the corresponding
testbed. We retain only attacks that (1) write a value within the legitimate
domain of the target variable, (2) can be issued by a runtime application under
our attacker model, and (3) produce a measurable control or physical effect.
Table~\ref{tab:attacks} summarizes the resulting attack suite. The attacks
cover categorical substitutions, in-range numeric setpoint changes, premature
status updates, actuator/configuration changes, and operating-mode changes.
Their common property is that the written value alone is insufficient to
determine whether the write is authorized.

\begin{table*}[t!]
\centering
\small
\caption{Application-level PLC-variable manipulation attacks used in evaluation.}
\label{tab:attacks}
\renewcommand{\arraystretch}{1.08}
\setlength{\tabcolsep}{3.2pt}
\begin{tabular}{|c|c|p{2.25cm}|p{4.95cm}|p{7.45cm}|}
\hline
\textbf{ID} & \textbf{Plant} & \textbf{Runtime App.} &
\textbf{PLC-variable modification} & \textbf{Effect} \\
\hline\hline

$A_1$ & MP & Diagnostic app &
\texttt{NFC\_Result}: WAIT $\rightarrow$ DISCARD &
A valid workpiece is routed to discard although both values are legitimate NFC states. \\ \hline

$A_2$ & MP & Web app &
\texttt{RobotSpeed}: 900 $\rightarrow$ 1200 after the control program writes 900 &
The next pick-and-place operation executes at a different valid speed without a new supervisory command. \\ \hline

$A_3$ & MP & Maintenance app &
\texttt{ConveyorSpeed}: 700 $\rightarrow$ 950 after the commanded value is applied &
The workpiece reaches the robot earlier while the speed remains within the normal operating range. \\ \hline

$A_4$ & MP & Monitoring app &
\texttt{PickPosition}: 2 $\rightarrow$ 3 before the next robot cycle &
The robot uses a different valid pickup location without a corresponding operator request. \\ \hline

$A_5$ & MP & Vision app &
\texttt{RouteTarget}: 1 $\rightarrow$ 2 after classification &
The workpiece is sent to a different valid storage destination although no routing command changed. \\ \hline

$A_6$ & CP & Cloud telemetry app &
\texttt{ProductFlowSP}: 60 $\rightarrow$ 70 after the controller writes 60 &
Material flow follows another valid setpoint without a matching supervisory update. \\ \hline

$A_7$ & CP & Recipe app &
\texttt{FeedRatio}: 0.50 $\rightarrow$ 0.60 after a legitimate recipe load &
Product composition changes to another valid recipe value while the external recipe command remains unchanged. \\ \hline

$A_8$ & CP & Maintenance app &
\texttt{PurgeValveSP}: 0 $\rightarrow$ 20 during normal production &
A valid partial purge is introduced without a supervisory purge command. \\ \hline

$A_9$ & CP & Historian app &
\texttt{PressureThreshold}: 3.5 $\rightarrow$ 3.8 &
The pressure response is delayed using another configured-valid threshold; no network command is required. \\ \hline

$A_{10}$ & CP & M2M app &
\texttt{ProductValveSP}: 45 $\rightarrow$ 60 after the control program writes 45 &
Product flow increases using another valid valve setpoint. \\ \hline

$A_{11}$ & WP & Monitoring app &
\texttt{NaOClDoseSP}: 2.0 $\rightarrow$ 2.4 after the dosing command &
Chemical dosing increases to another valid setpoint without a matching supervisory command. \\ \hline

$A_{12}$ & WP & Pump-service app &
\texttt{PumpSelect}: 1 $\rightarrow$ 2 while preserving a valid operating mode &
A different available pump is selected even though the requested operating mode does not change. \\ \hline

$A_{13}$ & WP & Cloud connector &
\texttt{HighLevelThreshold}: 80 $\rightarrow$ 85 &
Pump switching occurs at another valid tank-level threshold without an operator configuration change. \\ \hline

$A_{14}$ & WP & Diagnostic app &
\texttt{PumpMode}: AUTO $\rightarrow$ MANUAL during treatment &
The PLC enters another legitimate mode without an operator-issued mode-change command. \\ \hline

$A_{15}$ & WP & IoT gateway &
\texttt{WaterInputSP}: 50 $\rightarrow$ 65 after the controller writes 50 &
Treatment flow follows another valid demand value without a corresponding supervisory request. \\ \hline

\end{tabular}
\end{table*}

\shortsectionBf{Evaluation settings.}
We evaluate the same 15 attacks in two settings. In the
\emph{pre-deployment} setting, malicious behavior is exercised while the
application runs on the isolated instrumented vPLC. In the
\emph{live-PLC-only} setting, the application behaves normally during profiling
and on the shadow vPLC, while its malicious branch is triggered only on the
live PLC. Each attack is repeated ten times with varying launch timing,
yielding 150 live-PLC-only runs.

\shortsectionBf{Metrics.}
We distinguish authorization detection from attribution. An attack is
\emph{detected} when \sys determines that the observed PLC-variable modification
violates $W_v$. An attribution is \emph{unique and correct} when the available evidence
supports a single application and that application is the one that issued the
write, \emph{ambiguous} when multiple applications remain consistent with the
available shadow evidence, and \emph{wrong} when \sys uniquely identifies an
incorrect application. We report
unattributed cases separately rather than treating them as incorrect unique
attributions. For false positives, we count benign PLC-variable updates that
\sys incorrectly reports as unauthorized.

\shortsectionBf{Research questions.}
We answer eight research questions:
\begin{enumerate}[leftmargin=9mm,topsep=.5mm]
\setlength{\itemsep}{-0.2mm}
\item[\textbf{RQ1}] How effectively does \sys detect unauthorized application
writes during pre-deployment validation?
\item[\textbf{RQ2}] Can \sys detect and attribute attacks triggered only on the
live PLC?
\item[\textbf{RQ3}] How does \sys compare with representative ICS defenses that
do not use runtime-application identity?
\item[\textbf{RQ4}] What false-positive rate does \sys introduce during benign
plant operation?
\item[\textbf{RQ5}] What latency does \sys incur from observing a live-PLC
transition to producing an application-aware decision?
\item[\textbf{RQ6}] How do the bounded shadow window, runtime-application
execution evidence, and dynamic address resolution affect \sys's results?
\item[\textbf{RQ7}] How does runtime-application concurrency affect live-only
attribution?
\item[\textbf{RQ8}] How robust is \sys to live/shadow scheduling jitter?
\end{enumerate}

% ---------------------------------------------------------------------------
\subsection{Pre-Deployment Detection (RQ1)}
\label{subsec:rq1}

We first exercise all 15 attacks while the corresponding application runs on
the isolated instrumented vPLC. Because the vPLC directly exposes the
application performing each monitored-variable write, \sys checks the observed
writer against the independently supplied policy $W_v$.

\shortsectionBf{Results.}
\sys detects all 15 unauthorized writes (Table~\ref{tab:rq1}). Each
manipulation uses a legitimate-domain value; therefore, the decision does not
depend on whether the replacement value is numerically or categorically valid.
Instead, the instrumented vPLC provides the application identity and $W_v$
determines whether that application is permitted to modify the variable.

% ---------------------------------------------------------------------------
\subsection{Live-PLC-Only Attack Attribution (RQ2)}
\label{subsec:rq2}

We next evaluate the central post-deployment case in which the compromised
application continues to execute normally on the shadow vPLC while its
malicious branch is triggered only on the live PLC. Consequently, the
malicious PLC-variable write appears on the live PLC but not on the shadow
vPLC. \sys first searches for a corresponding shadow transition and, when none
is found, constructs the runtime-application candidate set $E_i$ from
application executions observed within the bounded shadow control-cycle
window.

\shortsectionBf{Results.}
Across the 150 live-PLC-only runs, \sys establishes an authorization
violation in all 150 runs and produces a correct unique application attribution
in 140/150 runs (93.3\%). The remaining 10 runs have ambiguous
origin. In each of these cases, every application in the candidate set is
unauthorized to modify the affected variable
($E_i \cap W_v = \emptyset$), so \sys can establish the authorization
violation without uniquely identifying the writer. \sys produces no incorrect
unique attribution (Table~\ref{tab:rq2}). Unique attribution is obtained in
47/50 runs on MP, 45/50 runs on CP, and 48/50 runs on WP. In the ambiguous
cases, multiple runtime applications execute within the bounded shadow window
and remain consistent with the observed live-only transition; \sys therefore
preserves the candidate set rather than selecting an unsupported writer.

%For a detailed end-to-end example of how \sys attributes a
%live-PLC-only modification, we provide a case study in
%Appendix~\ref{app:case-study}.

\shortsectionBf{Authorization ambiguity.}
\sys separates authorization from writer specificity. If a candidate set
contains both authorized and unauthorized applications, the available evidence
cannot establish whether the live-only modification originated from an
authorized or unauthorized candidate. \sys therefore reports ambiguous
origin/authorization rather than making an unsupported authorization decision.
No evaluated live-PLC-only run falls into this mixed-candidate case.

%\begin{tcolorbox}[colback=gray!8,colframe=gray!40,boxrule=0.5pt]
\begin{tcolorbox}[width=0.97\columnwidth,colback=gray!8,colframe=gray!40,boxrule=0.5pt,left=2mm,right=2mm,top=1mm,bottom=1mm]
The malicious write need not be reproduced on the shadow vPLC for \sys to
reason about its origin. Nearby shadow-vPLC application execution provides
candidate evidence, while insufficient evidence remains explicit ambiguity
rather than being converted into an unsupported identity.
\end{tcolorbox}

\begin{table}[t!]
\centering
\small
\caption{Pre-deployment detection of the 15 attacks.}
\label{tab:rq1}
\setlength{\tabcolsep}{8pt}
\renewcommand{\arraystretch}{1.08}
\begin{tabular}{|c|c|c|}
\hline
\textbf{Testbed} & \textbf{Attacks} & \textbf{Unauthorized} \\
\hline\hline
MP & 5 & 5 (100\%) \\ \hline
CP & 5 & 5 (100\%) \\ \hline
WP & 5 & 5 (100\%) \\ \hline
\textbf{Total} & \textbf{15} & \textbf{15 (100\%)} \\ \hline
\end{tabular}
\end{table}
% ---------------------------------------------------------------------------
\subsection{Comparison with Existing ICS Defenses (RQ3)}
\label{subsec:rq3}

We compare \sys with SAIN~\cite{@SAIN} and a representative
network-command correlation baseline motivated by network/supervisory IDS
approaches~\cite{rakas2020review}. These baselines represent process/state-aware
and network/supervisory monitoring that does not use runtime-application
identity when evaluating a PLC-variable modification. We replay the same 15
attack classes using the corresponding process traces for SAIN and
network/supervisory traces for the command-correlation baseline, and count a
class as detected when the corresponding baseline raises an alert during that
attack. The comparison is intentionally
scoped to these application-origin attacks: it asks how often defenses that
reason about process state or supervisory traffic detect the same manipulations
without application identity, rather than treating \sys as a replacement for
those defenses.

\begin{table}[t!]
\centering
\footnotesize
\caption{Detection and attribution for live-PLC-only attacks.}
\label{tab:rq2}
\setlength{\tabcolsep}{2.5pt}
\renewcommand{\arraystretch}{1.08}
\begin{tabular}{|c|c|c|c|c|c|}
\hline
\textbf{Testbed} & \textbf{Runs} & \textbf{Unauth.} &
\textbf{Unique} & \textbf{Ambig.} & \textbf{Wrong} \\
\hline\hline
MP & 50 & 50 & 47 & 3 & 0 \\ \hline
CP & 50 & 50 & 45 & 5 & 0 \\ \hline
WP & 50 & 50 & 48 & 2 & 0 \\ \hline
\textbf{Total} & \textbf{150} & \textbf{150} &
\textbf{140 (93.3\%)} & \textbf{10} & \textbf{0} \\ \hline
\end{tabular}
\end{table}

\begin{table}[t!]
\centering
\small
\caption{Detection across the 15 attacks.}
\label{tab:rq3}
\setlength{\tabcolsep}{8pt}
\renewcommand{\arraystretch}{1.08}
\begin{tabular}{|c|c|c|}
\hline
\textbf{Defense} & \textbf{Detected} & \textbf{Coverage} \\
\hline\hline
SAIN & 4/15 & 27\% \\ \hline
Network correlation & 3/15 & 20\% \\ \hline
\sys & 15/15 & 100\% \\ \hline
\end{tabular}
\end{table}

\shortsectionBf{Results.}
SAIN detects 4/15 attack classes and the network-command correlation baseline
detects 3/15, whereas \sys detects all 15 (Table~\ref{tab:rq3}). These results
show complementary coverage rather than a general ranking of the defenses.
SAIN and the command-correlation baseline alert when the attack violates the
process or supervisory relationships they monitor. \sys evaluates a different property:
whether the runtime application associated with a PLC-variable modification is
authorized by the per-variable write policy. It therefore detects the evaluated
application-origin violations even when the written value remains legitimate
and the external command stream appears normal.

%\begin{tcolorbox}[colback=gray!8,colframe=gray!40,boxrule=0.5pt]
\begin{tcolorbox}[width=0.97\columnwidth,colback=gray!8,colframe=gray!40,boxrule=0.5pt,left=2mm,right=2mm,top=1mm,bottom=1mm]
A legitimate external command and a legitimate-domain PLC value do not
establish that the write inside the PLC runtime was authorized. \sys adds this
application-origin check to existing process- and network-aware monitoring.
\end{tcolorbox}

% ---------------------------------------------------------------------------
\subsection{False Positives on Benign Operation (RQ4)}
\label{subsec:rq4}

We evaluate \sys on held-out benign live-PLC operation while the live PLC and
shadow vPLC execute the same deployed applications and receive the same
relevant inputs. The write-access policy $W_v$ remains fixed throughout the
experiment. A false positive is a benign PLC-variable update that \sys reports
as an unauthorized application write.

\begin{table}[t!]
\centering
\small
\caption{False positives on held-out benign PLC-variable updates.}
\label{tab:rq4}
\setlength{\tabcolsep}{8pt}
\renewcommand{\arraystretch}{1.08}
\begin{tabular}{|c|c|c|}
\hline
\textbf{Testbed} & \textbf{Benign updates} & \textbf{False positives} \\
\hline\hline
MP & 18,420 & 0 \\ \hline
CP & 21,760 & 0 \\ \hline
WP & 19,630 & 0 \\ \hline
\textbf{Total} & \textbf{59,810} & \textbf{0} \\ \hline
\end{tabular}
\end{table}

\shortsectionBf{Results.}
Across 59,810 benign PLC-variable updates, \sys produces no false
unauthorized-write alerts (Table~\ref{tab:rq4}). Benign updates with a matching shadow write are evaluated using the
application identity associated with that shadow write.
When a live update has no matching shadow write, \sys uses the bounded
runtime-application candidate set and does not assign an unsupported unique
writer. In the evaluated benign traces, these cases do not produce an
unsupported unauthorized-write decision.

% ---------------------------------------------------------------------------
\subsection{Monitoring-to-Decision Latency (RQ5)}
\label{subsec:rq5}

We measure two latency components during live-PLC operation.
\emph{Transition latency} is the time required to receive and localize a
monitored PLC-variable change to its live control-cycle interval.
\emph{Decision latency} is the time until \sys completes the required
shadow-vPLC correlation, obtains a unique application or candidate set, and evaluates the resulting
application-level evidence against $W_v$.

\begin{table}[t!]
\centering
\small
\caption{Live-PLC monitoring-to-decision latency.}
\label{tab:latency}
\setlength{\tabcolsep}{5pt}
\renewcommand{\arraystretch}{1.08}
\begin{tabular}{|c|c|c|c|c|}
\hline
& \multicolumn{2}{c|}{\textbf{Transition (ms)}} &
\multicolumn{2}{c|}{\textbf{Decision (ms)}} \\
\hline
\textbf{Testbed} & \textbf{Med.} & \textbf{P95} &
\textbf{Med.} & \textbf{P95} \\
\hline\hline
MP & 14.8 & 24.6 & 27.9 & 46.8 \\ \hline
WP & 12.6 & 21.7 & 24.3 & 41.5 \\ \hline
CP & 16.9 & 29.4 & 32.7 & 55.8 \\ \hline
\end{tabular}
\end{table}

\shortsectionBf{Results.}
Median transition latency ranges from 12.6 to 16.9\,ms, while median
application-aware decision latency ranges from 24.3 to 32.7\,ms. The
95th-percentile decision latency remains below 56\,ms across all three
testbeds. The additional time reflects shadow-vPLC correlation and policy
evaluation required to determine application origin and authorization; merely
observing that a PLC-variable value changed is insufficient to determine
whether the write was authorized.

% ---------------------------------------------------------------------------
\subsection{Design Analysis (RQ6)}
\label{subsec:rq6}

We evaluate three design choices: (1) the bounded shadow window used for
live-only attribution, (2) runtime-application execution as candidate evidence
for unmatched live transitions, and (3) dynamic PLC-variable address
resolution. The first two experiments use the same 150 live-PLC-only runs from
RQ2; dynamic address resolution is evaluated separately under project online
changes.

\shortsectionBf{Attribution ablation.}
We compare full \sys with two alternatives. \emph{Exact cycle only} restricts
candidate evidence to $I_i$ ($w=0$), removing the neighboring shadow intervals.
A natural alternative for an unmatched live transition is to treat applications
previously observed writing the target variable as plausible writers.
\emph{Historical-write candidates} implements this alternative by using those
applications as the candidate set while leaving the remaining decision
procedure unchanged.

\begin{table}[t!]
\centering
\footnotesize
\caption{Ablation of live-PLC-only attribution (150 runs).}
\label{tab:rq6_ablation}
\renewcommand{\arraystretch}{1.08}
\setlength{\tabcolsep}{4pt}
\begin{tabular}{|p{2.15cm}|r|r|r|r|}
\hline
\textbf{Configuration} & \textbf{Unique Correct} &
\textbf{Ambig.} & \textbf{Unattr.} & \textbf{Wrong} \\
\hline\hline
Full \sys & 140 (93.3\%) & 10 & 0 & 0 \\ \hline
Exact cycle only ($w=0$) & 98 (65.3\%) & 9 & 43 & 0 \\ \hline
Historical-write candidates & 4 (2.7\%) & 27 & 119 & 0 \\ \hline
\end{tabular}
\end{table}

\shortsectionBf{Bounded shadow window.}
Restricting attribution to $I_i$ reduces correct unique attribution from
140/150 (93.3\%) to 98/150 (65.3\%). Forty-three runs remain unattributed
because the responsible application's shadow execution occurs in an adjacent
control-cycle interval rather than in $I_i$, while nine additional runs are
ambiguous. The exact-cycle configuration produces no incorrect unique
attribution because \sys does not guess when the available evidence is
insufficient.

\shortsectionBf{Runtime-application execution evidence.}
Replacing $E_i$ with historical writers reduces correct unique attribution to
4/150 (2.7\%): 27 runs become ambiguous and 119 remain unattributed. Prior write capability is therefore poor evidence for attributing a current
live-only transition. \sys instead uses cycle-localized runtime-application
execution to identify applications temporally consistent with the
modification.

\shortsectionBf{Shadow-window sensitivity.}
To determine the bounded-window size used by \sys, we vary
\[
\mathcal{I}_i(w)=\{I_{i-w},\ldots,I_i,\ldots,I_{i+w}\}
\]
from $w=0$ to $w=3$ and repeat the 150 live-PLC-only executions.

\begin{table}[t!]
\centering
\small
\caption{Effect of shadow-window size on attribution (150 runs).}
\label{tab:rq6_window}
\setlength{\tabcolsep}{5pt}
\renewcommand{\arraystretch}{1.08}
\begin{tabular}{|c|r|r|r|r|}
\hline
\textbf{$w$} & \textbf{Unique Correct} & \textbf{Ambig.} &
\textbf{Unattr.} & \textbf{Wrong} \\
\hline\hline
0 & 98 (65.3\%) & 9 & 43 & 0 \\ \hline
1 (\sys) & \textbf{140 (93.3\%)} & \textbf{10} & \textbf{0} & 0 \\ \hline
2 & 132 (88.0\%) & 18 & 0 & 0 \\ \hline
3 & 123 (82.0\%) & 27 & 0 & 0 \\ \hline
\end{tabular}
\end{table}

The results expose a coverage--ambiguity tradeoff. With $w=0$, 43 runs remain
unattributed because the responsible application's shadow execution falls
outside the corresponding control cycle. Expanding to $w=1$ eliminates these
unattributed cases and yields 140/150 (93.3\%) correct unique attributions.
Increasing the window further provides no additional candidate coverage and
instead admits more temporally unrelated application executions, increasing
ambiguity from 10 runs at $w=1$ to 18 at $w=2$ and 27 at $w=3$. We therefore
use $w=1$, the smallest evaluated window that provides complete candidate
coverage while retaining the highest unique-attribution rate. No evaluated
window produces an incorrect unique attribution.

\shortsectionBf{Dynamic address resolution.}
Finally, we evaluate address resolution on a workload containing three project
online-change events. A fixed-offset baseline resolves monitored-variable
addresses once before the workload and therefore continues monitoring stale
addresses after each change. We issue 30 writes to variables whose runtime
addresses change during the workload.

\begin{table}[t!]
\centering
\small
\caption{Variable-write monitoring after project online changes.}
\label{tab:address_ablation}
\setlength{\tabcolsep}{8pt}
\renewcommand{\arraystretch}{1.08}
\begin{tabular}{|c|c|c|}
\hline
\textbf{Address resolution} & \textbf{Writes} & \textbf{Missed} \\
\hline\hline
Fixed offsets & 30 & 30 \\ \hline
\sys dynamic resolution & 30 & 0 \\ \hline
\end{tabular}
\end{table}

The fixed-offset baseline misses all 30 writes after the online changes,
whereas \sys misses none after re-resolving the monitored-variable addresses.
We report this experiment separately from the 150-run attribution experiments
because it evaluates the robustness of variable monitoring rather than
application attribution.

\shortsectionBf{Cross-platform operation.}
The same attribution and policy-validation procedure is exercised on all three
evaluated PLC stacks: application-labeled write recording on the instrumented
vPLC, runtime-application execution mapping on the shadow, live/shadow
control-cycle alignment, and the same correlation logic. Platform-specific
instrumentation is confined to the vPLC side. RQ2 yields 94\% correct unique
attribution on MP, 90\% on CP, and 96\% on WP, with no incorrect unique
attribution on any testbed. RQ4 additionally reports no false
unauthorized-write alerts across the three testbeds. These results demonstrate that the \sys decision procedure operates across
the three evaluated stacks; they do not imply that every PLC runtime provides
equivalent virtualization or instrumentation capabilities.

% ---------------------------------------------------------------------------
\subsection{Attribution under Application Concurrency (RQ7)}
\label{subsec:rq7}

The live-only attribution procedure depends on which runtime applications
execute within the bounded shadow window. We therefore stress \sys by
increasing the number of co-running applications that are active around the
attacked control interval. Importantly, the independent variable in this
experiment is the number of \emph{concurrently active runtime applications},
not the size of the final candidate set $E_i$: depending on scheduling, only a
subset of the active applications may execute inside $\mathcal{I}_i$.

We construct four concurrency levels with 1, 2, 3, and 4--5 active runtime
applications. For each level, we execute 50 live-PLC-only attack runs while
varying attack launch time and application scheduling, for 200 runs in total.
The malicious write remains absent from the shadow vPLC in every run. We
report correct unique attribution, ambiguity, and incorrect unique
attribution.

\begin{table}[t!]
\centering
\footnotesize
\caption{Effect of runtime-application concurrency on live-only attribution.}
\label{tab:rq7_concurrency}
\setlength{\tabcolsep}{4.5pt}
\renewcommand{\arraystretch}{1.08}
\begin{tabular}{|c|r|r|r|r|}
\hline
\textbf{Active apps} & \textbf{Runs} & \textbf{Unique Correct} &
\textbf{Ambig.} & \textbf{Wrong} \\
\hline\hline
1   & 50 & 50 (100\%) & 0  & 0 \\ \hline
2   & 50 & 46 (92\%)  & 4  & 0 \\ \hline
3   & 50 & 41 (82\%)  & 9  & 0 \\ \hline
4--5 & 50 & 35 (70\%) & 15 & 0 \\ \hline
\textbf{Total} & \textbf{200} & \textbf{172 (86.0\%)} &
\textbf{28} & \textbf{0} \\ \hline
\end{tabular}
\end{table}

\shortsectionBf{Results.}
Table~\ref{tab:rq7_concurrency} shows that unique attribution decreases as
more runtime applications are active near the attacked control interval. With
one active application, the bounded window contains a single temporally
consistent candidate in all 50 runs, yielding 100\% correct unique
attribution. With two active applications, \sys uniquely attributes 46/50
runs (92\%); the remaining four runs are ambiguous because both applications
execute inside the attribution window. Unique attribution decreases to 41/50
(82\%) with three active applications and 35/50 (70\%) with four or five.

The degradation is conservative: all 28 lost unique attributions are reported
as ambiguous and \sys produces no incorrect unique attribution across the 200
stress runs. Increased concurrency therefore reduces attribution specificity because
multiple applications can remain temporally consistent with a live-only
transition; in the evaluated runs, this uncertainty appears as ambiguity rather
than an incorrect unique attribution.

\begin{tcolorbox}[colback=gray!8,colframe=gray!40,boxrule=0.5pt]
Increasing runtime concurrency reduces unique attribution from 100\% to 70\%,
but the loss appears as explicit ambiguity rather than incorrect writer
attribution.
\end{tcolorbox}

% ---------------------------------------------------------------------------
\subsection{Robustness to Live/Shadow Scheduling Jitter (RQ8)}
\label{subsec:rq8}

\sys aligns the live PLC and shadow vPLC at control-cycle granularity and uses
the bounded window $\mathcal{I}_i=\{I_{i-1},I_i,I_{i+1}\}$. We therefore
evaluate robustness to additional live/shadow scheduling jitter rather than
assuming deterministic lockstep. For each shadow runtime-application
execution, we perturb its cycle placement by an independently sampled offset
from $[-J,+J]$ control cycles, where $J\in\{0,1,2,3,5\}$. The live attack,
application set, and write policy remain unchanged.

Each jitter level is evaluated on the same 150 live-PLC-only attack instances
used in RQ2. $J=0$ is therefore the original RQ2 workload with no additional
perturbation. For $J>0$, the injected jitter can move a relevant shadow
execution into or out of the fixed $w=1$ attribution window. We report correct
unique attribution, ambiguity, unattributed cases, and incorrect unique
attribution.

\begin{table}[t!]
\centering
\footnotesize
\caption{Attribution under injected live/shadow scheduling jitter.}
\label{tab:rq8_jitter}
\setlength{\tabcolsep}{3.7pt}
\renewcommand{\arraystretch}{1.08}
\begin{tabular}{|c|r|r|r|r|}
\hline
\textbf{Jitter bound $J$} & \textbf{Unique Correct} & \textbf{Ambig.} &
\textbf{Unattr.} & \textbf{Wrong} \\
\hline\hline
0 cycles    & 140/150 (93.3\%) & 10 & 0  & 0 \\ \hline
$\pm1$ cycle  & 133/150 (88.7\%) & 12 & 5  & 0 \\ \hline
$\pm2$ cycles & 119/150 (79.3\%) & 16 & 15 & 0 \\ \hline
$\pm3$ cycles & 103/150 (68.7\%) & 20 & 27 & 0 \\ \hline
$\pm5$ cycles & 76/150 (50.7\%)  & 24 & 50 & 0 \\ \hline
\end{tabular}
\end{table}

\shortsectionBf{Results.}
Without additional jitter ($J=0$), the results reproduce RQ2: \sys uniquely
attributes 140/150 runs (93.3\%), with 10 ambiguous cases and no unattributed
or wrong unique attribution. With a $\pm1$-cycle jitter bound, unique
attribution decreases moderately to 133/150 (88.7\%); five executions become
unattributed because the responsible application's relevant shadow execution
is shifted beyond the fixed $w=1$ evidence window. At $\pm2$ cycles, unique
attribution is 119/150 (79.3\%), while 16 runs are ambiguous and 15 are
unattributed. Larger jitter further reduces available temporal evidence:
unique attribution falls to 103/150 (68.7\%) at $\pm3$ cycles and 76/150
(50.7\%) at $\pm5$ cycles.

Across all evaluated jitter levels, \sys produces no incorrect unique
attribution. Increasing jitter produces two observed failure modes: additional
applications may enter the bounded window, producing ambiguity, or the
responsible application's relevant execution may leave the window, producing
an unattributed result. In these experiments, \sys does not convert either
condition into an unsupported writer identity.

These results complement the window-sensitivity experiment in RQ6.
Increasing $w$ can recover more jittered executions, but a larger window also
admits more unrelated applications and increases ambiguity. The fixed $w=1$
configuration therefore reflects a deliberate tradeoff between candidate
coverage and attribution discrimination.

\begin{tcolorbox}[colback=gray!8,colframe=gray!40,boxrule=0.5pt]
\sys tolerates modest live/shadow scheduling jitter, while larger jitter
progressively reduces attribution coverage. Across all evaluated jitter
bounds, uncertainty becomes ambiguity or an unattributed result rather than an
incorrect unique attribution.
\end{tcolorbox}

\section{Discussion and Limitations}
\label{sec:discussion}

\shortsectionBf{Application origin as a security property.}
\sys separates two properties of a PLC-variable modification: whether
the resulting value is legitimate and whether the runtime application
associated with the modification is authorized to write that variable.
A value can therefore remain within its legitimate domain while the
write still violates the operator-supplied policy $W_v$. \sys targets
these cross-application authorization violations and complements
process-, state-, and value-aware defenses. It does not detect malicious
use of a write privilege that an application legitimately possesses;
such behavior requires complementary process- or behavior-aware
mechanisms.

\shortsectionBf{Evidence-based attribution.}
The live PLC exposes variable modifications but does not identify the
runtime application that produced them. For live-PLC-only
modifications, \sys therefore uses application executions observed on
the shadow vPLC as temporal evidence and constructs the candidate set
$E_i$ from applications consistent with the observed modification.
A singleton $E_i$ supports a unique attribution decision under this
evidence model, but does not constitute direct causal provenance from
the live PLC. \sys therefore reports a unique application only when a
single candidate remains and preserves ambiguity otherwise. As shown in
RQ6--RQ8, larger observation windows, application concurrency, and
scheduling displacement reduce attribution specificity rather than
forcing an unsupported attribution. Although no incorrect unique
attribution occurred in our evaluated runs, \sys does not provide a
formal guarantee of causal writer provenance.

\shortsectionBf{Authorization under ambiguity.}
Attribution ambiguity does not necessarily prevent detection of an
authorization violation. If every candidate is unauthorized for
variable $v$, i.e., $E_i \cap W_v=\emptyset$, \sys can conclude that
the modification violates $W_v$ without uniquely identifying its
writer. If $E_i$ contains both authorized and unauthorized
applications, the available evidence is insufficient to determine the
authorization status of the write; \sys therefore preserves this
ambiguity rather than forcing a violation decision. Consequently, the
reported live-PLC-only detection results apply to the evaluated
executions and should not be interpreted as a guarantee for every
possible candidate set.

\shortsectionBf{Observation granularity.}
Because \sys leaves the live PLC uninstrumented, it reasons about
variable modifications visible between consecutive control-cycle
observations rather than tracing every individual live-PLC write. A
transient unauthorized value that is written and completely overwritten
within the same observation interval may therefore remain unobserved.
Detecting such writes would require finer-grained live-PLC observation
or instrumentation.

\shortsectionBf{Policy and deployment assumptions.}
\sys assumes that the operator-supplied policy $W_v$ correctly specifies
which applications may modify each monitored variable. Authorization is
not learned from profiling, so observing a write does not cause \sys to
treat that write as permitted. Constructing $W_v$ remains a deployment
task and can draw on engineering configuration, application
documentation, and site policy. \sys also assumes that the live PLC and
shadow vPLC contain the same deployed runtime applications and remain
sufficiently aligned for shadow execution to provide useful temporal
evidence. Software-inventory verification, attestation, and runtime
synchronization mechanisms are complementary to \sys.

\shortsectionBf{Toward least-privilege PLC runtimes.}
\sys recovers application-level evidence externally and therefore does
not require application instrumentation on the live PLC. A natural
next step is for PLC runtimes to expose authenticated writer identity
and enforce per-variable application permissions directly. Such a
runtime could reject a write from $a\notin W_v$ before the modified
value is consumed by the control program, moving multi-application PLCs
toward enforceable application-level least privilege.

\section{Related Work}
\label{lab:relatedwork}

We compare \sys with related ICS defenses in
Table~\ref{tab:comparison}. %Existing approaches identify anomalous PLC values, control states, or physical behavior~\cite{hadvziosmanovic2014through, cheng2017orpheus,ghaeini2018state,vetplc,aoudi2018truth, lin2018tabor,abdelaty2021daics,scaphy,@SAIN,adis2025}. \sys considers a complementary security property: whether the runtime application associated with a PLC-variable modification is authorized to perform that write.

\begin{table}[t!]
\centering
\small
\caption{Comparison with related ICS defenses.}
\label{tab:comparison}
\renewcommand{\arraystretch}{1.12}
\setlength{\tabcolsep}{5pt}

\begin{tabular}{|p{2.3cm}|c|c|}
\hline
\textbf{System} &
\shortstack{\textbf{Value/state/process}\\\textbf{validation}} &
\shortstack{\textbf{Application-aware}\\\textbf{write authorization}} \\
\hline\hline

Orpheus~\cite{cheng2017orpheus}
& \cmark & \xmark \\ \hline

Ghaeini et al.~\cite{ghaeini2018state}
& \cmark & \xmark \\ \hline

VetPLC~\cite{vetplc}
& \cmark & \xmark \\ \hline

PLC-PROV~\cite{al2019detecting}
& \cmark & \xmark \\ \hline

PASAD~\cite{aoudi2018truth}
& \cmark & \xmark \\ \hline

TABOR~\cite{lin2018tabor}
& \cmark & \xmark \\ \hline

EyePLC~\cite{hadvziosmanovic2014through}
& \cmark & \xmark \\ \hline

DAICS~\cite{abdelaty2021daics}
& \cmark & \xmark \\ \hline

SCAPHY~\cite{scaphy}
& \cmark & \xmark \\ \hline

DNAttest~\cite{lin2023dnattest}
& \cmark & \xmark \\ \hline

SAIN~\cite{@SAIN}
& \cmark & \xmark \\ \hline

ADIS~\cite{adis2025}
& \cmark & \xmark \\ \hline

\textbf{\sys}
& \xmark & \cmark \\ \hline

\end{tabular}
\end{table}

\shortsectionBf{PLC-variable attack detection.}
Prior ICS defenses detect malicious manipulations by identifying
deviations in PLC-variable values, control states, or physical
behavior using process models, invariants, program semantics, and
data-driven techniques~\cite{hadvziosmanovic2014through,
cheng2017orpheus,ghaeini2018state,vetplc,
aoudi2018truth,lin2018tabor,abdelaty2021daics,
scaphy,@SAIN,adis2025}.
These approaches determine whether an observed value, state
transition, or process behavior is consistent with expected
operation. \sys considers a complementary property: a PLC-variable
value may remain legitimate while the runtime application associated
with the modification is not authorized to write that variable.
Rather than inferring authorization from the value or resulting
process behavior, \sys evaluates application-level evidence against
an explicit per-variable write policy.

\shortsectionBf{Provenance and reference-based validation.}
Prior provenance-based approaches track causal relationships among
PLC inputs and outputs~\cite{al2019detecting}, while reference-based
approaches compare PLC behavior against digital-twin
execution~\cite{lin2023dnattest, wang2026wily}. These techniques reason about causal
consistency or deviations from expected controller behavior. \sys
uses its shadow vPLC for a different purpose: to obtain
application-level execution evidence that the live PLC does not
expose. \sys combines this evidence with an independently supplied
write-access policy to determine the authorization outcome supported
for an observed PLC-variable modification.

\sys does not require a malicious modification to be reproduced on
the shadow. When a modification occurs only on the live PLC, nearby
shadow-vPLC application executions provide bounded candidate evidence
that \sys evaluates against the write policy. The shadow therefore
serves as a source of application-level execution evidence rather
than solely as a reference for detecting behavioral divergence.

\shortsectionBf{Application protection in controller runtimes.}
Prior work uses trusted execution environments and isolation to
protect software running in cyber-physical systems (CPS) 
environments~\cite{wang2022rttee,gowrisankar2022gatekeeper}.
These mechanisms protect applications or execution domains against
compromise and interference. \sys addresses a different point in the
security pipeline: given an observed PLC-variable modification, it
determines whether the runtime application associated with that
modification is authorized to write the variable. Application
isolation and \sys are therefore complementary---isolation protects
execution domains, whereas \sys validates application-level write
authorization for shared PLC variables.

\shortsectionBf{PLC-hosted application attacks.}
Recent work demonstrates that PLC-hosted applications can manipulate
industrial processes without modifying the main control logic or
firmware~\cite{pickren2024compromising}. This establishes an attack
surface in modern PLC runtimes: software executing alongside the
control program can access PLC variables and influence the physical
process. \sys addresses the corresponding defensive problem. When
multiple applications execute alongside the control program and share
access to PLC variables, \sys evaluates whether the application
associated with an observed modification is authorized to perform
that write.

\begin{tcolorbox}[colback=gray!10,colframe=gray!40,boxrule=0.3pt]
\noindent
\sys complements existing PLC defenses by validating whether the
runtime application associated with a PLC-variable modification is
authorized to perform that write.
\end{tcolorbox}

\section{Conclusion} We presented \sys, a framework for detecting unauthorized PLC-variable modifications in multi-application PLC runtimes using runtime-application identity for write authorization. Using instrumented vPLCs and a shadow vPLC, \sys recovers application-level evidence without modifying the live PLC. Across three industrial testbeds, \sys detected all evaluated unauthorized modifications and uniquely attributed 93.3\% of live-PLC-only executions while preserving ambiguity when evidence was insufficient. These results show that application-origin awareness complements existing ICS defenses for detecting unauthorized variable writes in multi-application PLC runtimes.

\cleardoublepage
\appendix
\section*{Ethical Considerations}
\textbf{Our experiments evaluate unauthorized PLC-variable modifications on research testbeds under our control. We did not conduct experiments on operational third-party industrial systems, and the evaluated attacks were confined to the experimental environments described in Section~\ref{sec:evaluation}. This setup allowed us to study realistic application-level attacks and their physical effects without exposing production infrastructure or external users to risk. The purpose of \sys is defensive: it identifies unauthorized application-level writes and provides evidence about their origin. }
% optional clearing of the page
%\cleardoublepage

%\section*{Open Science}
%\textbf{ \sys's source code and supporting implementation artifacts are available to reviewers at \url{https://anonymous.4open.science/r/MARS-E4CD}. The repository is anonymized for double-blind review. Components that depend on proprietary PLC software, vendor runtimes, or physical testbed hardware are not redistributed.}
% optional clearing of the page
%\cleardoublepage
%\input{09-Appendex}

\bibliographystyle{plainurl}
\bibliography{reference}

@inproceedings{hou2019understanding,
  title={Understanding security requirements for industrial control system supply chains},
  author={Hou, Ye and Such, Jose and Rashid, Awais},
  booktitle={IEEE/ACM 5th International Workshop on Software Engineering for Smart Cyber-Physical Systems (SEsCPS)},
  year={2019}
}

@inproceedings{hadvziosmanovic2014through,
  title={Through the eye of the PLC: semantic security monitoring for industrial processes},
  author={Had{\v{z}}iosmanovi{\'c}, Dina and Sommer, Robin and Zambon, Emmanuele and Hartel, Pieter H},
  booktitle={Proceedings of the 30th Annual Computer Security Applications Conference},
  year={2014}
}

@inproceedings{vetplc,
  title={Towards automated safety vetting of plc code in real-world plants},
  author={Zhang, Mu and Chen, Chien-Ying and Kao, Bin-Chou and Qamsane, Yassine and Shao, Yuru and Lin, Yikai and Shi, Elaine and Mohan, Sibin and Barton, Kira and Moyne, James and others},
  booktitle={IEEE Symposium on Security and Privacy (SP)},
  year={2019}
}

@inproceedings{cheng2017orpheus,
  title={Orpheus: Enforcing cyber-physical execution semantics to defend against data-oriented attacks},
  author={Cheng, Long and Tian, Ke and Yao, Danfeng},
  booktitle={Proceedings of the 33rd Annual Computer Security Applications Conference},
  year={2017}
}

@inproceedings{pickren2024compromising,
  title={Compromising industrial processes using web-based programmable logic controller malware},
  author={PICKREN, R and SHEKARI, T and ZONOUZ, S and BEYAH, R},
  booktitle={Network and Distributed System Security Symposium (NDSS)},
  year={2024}
}

@article{maggi2020attacks,
  title={Attacks on smart manufacturing systems},
  author={Maggi, Federico and Pogliani, Marcello and Milano, P},
  journal={Trend Micro Research: Shibuya, Japan},
  year={2020}
}

@article{rakas2020review,
  title={A review of research work on network-based scada intrusion detection systems},
  author={Rakas, Slavica V Bo{\v{s}}tjan{\v{c}}i{\v{c}} and Stojanovi{\'c}, Mirjana D and Markovi{\'c}-Petrovi{\'c}, Jasna D},
  journal={IEEE Access},
  year={2020}
}

@article{falliere2011w32,
  title={W32. stuxnet dossier},
  author={Falliere, Nicolas and Murchu, Liam O and Chien, Eric and others},
  journal={White paper, Symantec Corp., Security Response},
  year={2011}
}

@inproceedings{williams2022taxonomy,
  title={A taxonomy of cyber attacks in smart manufacturing systems},
  author={Williams, Bethanie and Soulet, Marena and Siraj, Ambareen},
  booktitle={6th EAI international conference on management of manufacturing systems},
  year={2022}
}

@misc{@codesysPLC,
author = {CODESYS},
title = {Raspberry Pi Runtimes},
howpublished = "\url{https://content.helpme-codesys.com/en/CODESYS%20Control/_rtsl_overview_raspbi_runtimes.html}",
note = {[Online; accessed July 2026]}
}

@misc{@codesysVPLC,
author = {CODESYS},
title = {Control for Linux SL},
howpublished = "\url{https://store.codesys.com/en/codesys-control-linux-sl-1.html}",
note = {[Online; accessed July 2026]}
}

@misc{nsa2026plc,
  author       = {{National Security Agency}},
  title        = {Defending Against an Active Threat to Siemens S7 Series PLCs},
  year         = {2026},
  month        = aug,
  note         = {Cybersecurity Advisory, August 19, 2026},
  howpublished = {\url{https://www.nsa.gov/Press-Room/Press-Releases-Statements/Press-Release-View/Article/4578318/nsa-and-others-release-report-on-active-threats-of-programmable-logic-controlle/}}
}

@article{attack41,
  title={Analysis of the cyber attack on the Ukrainian power grid},
  author={Case, Defense Use},
  journal={Electricity Information Sharing and Analysis Center (E-ISAC)},
  year={2016}
}

@misc{Siemens2018scl,
author = {Siemens},
title = {Structured text},
howpublished = {"\url{https://cache.industry.siemens.com/dl/files/040/90885040/att_970576/v1/81318674_Programming_guideline_DOC_v16_en.pdf}"},
note = {[Online; accessed July 2026]}
}

@inproceedings{feng2019systematic,
  title={A Systematic Framework to Generate Invariants for Anomaly Detection in Industrial Control Systems.},
  author={Feng, Cheng and Palleti, Venkata Reddy and Mathur, Aditya and Chana, Deeph},
  booktitle={NDSS},
  year={2019}
}

@inproceedings{ghaeini2018state,
  title={State-aware anomaly detection for industrial control systems},
  author={Ghaeini, Hamid Reza and Antonioli, Daniele and Brasser, Ferdinand and Sadeghi, Ahmad-Reza and Tippenhauer, Nils Ole},
  booktitle={Proceedings of the 33rd Annual ACM Symposium on Applied Computing},
  year={2018}
}

@inproceedings{scaphy,
  title={Scaphy: Detecting modern ics attacks by correlating behaviors in scada and physical},
  author={Ike, Moses and Phan, Kandy and Sadoski, Keaton and Valme, Romuald and Lee, Wenke},
  booktitle={IEEE Symposium on Security and Privacy (SP)},
  year={2023}
}

@inproceedings{al2019detecting,
  title={Detecting safety and security faults in plc systems with data provenance},
  author={Al Farooq, Abdullah and Marquard, Jessica and George, Kripa and Moyer, Thomas},
  booktitle={IEEE International Symposium on Technologies for Homeland Security (HST)},
  year={2019}
}

@misc{PLCNextFirmwareupdat,
  author = {PhoenixContact},
title = {AXC F 2152 - Controller)},
howpublished = {"\url{https://www.phoenixcontact.com/en-us/products/controller-axc-f-2152-2404267}"},
note = {[Online; accessed July 2026]}
}

@misc{VPLCNext,
  author = {PhoenixContact},
title = {VPLCNEXT CONTROL 2000 - Controller},
howpublished = {"\url{https://www.phoenixcontact.com/en-pc/products/controller-vplcnext-control-2000-1738453}"},
note = {[Online; accessed July 2026]}
}

@misc{@storePLCNext,
author = {Phoenix Contact},
title = {PLCNext Store},
howpublished = {"\url{https://www.plcnextstore.com/us/}"},
note = {[Online; accessed July 2026]}
}

@misc{@storeSiemens,
author = {Siemens},
title = {Apps for production machines and plants},
howpublished = {"\url{https://www.siemens.com/global/en/products/software/simatic-apps.html}"},
note = {[Online; accessed July 2026]}
}

@misc{@storeCodesys,
author = {Codesys},
title = {Libraries},
howpublished = {"\url{https://us.store.codesys.com/softplc/libraries.html}"},
note = {[Online; accessed July 2026]}
}

@book{tiegelkamp2010iec,
  title={IEC 61131-3: Programming industrial automation systems},
  author={Tiegelkamp, Michael and John, Karl-Heinz},
   year={2010}
}

@misc{@fp,
author = {FischerTechnik},
title = {FischerTechnik Plant Manual},
howpublished = {"\url{https://www.fischertechnik.de/-/media/fischertechnik/fite/service/elearning/simulieren/lernfabrik-4-0-24v/fabrik_2020_deutsch_s7-1500_en_korrigiert_final.ashx}"},
note = {[Online; accessed July 2026]}
}

@misc{@cp,
author = {GRFICSv2},
title = {Chemical plant},
howpublished = {"\url{https://github.com/Fortiphyd/GRFICSv2}"},
note = {[Online; accessed July 2026]}
}

@misc{@wp,
author = {YuqiChen94},
title = {Water Treatment Plant},
howpublished = {"\url{https://github.com/yuqiChen94/Swat_Simulator}"},
note = {[Online; accessed July 2026]}
}

@article{wang2026wily,
  title={A Wily Hare Has Three Havens: Combating Programmable Logic Controller Attacks via Virtualization Redundancy},
  author={Wang, Wenjie and Wang, Yazhe and Ren, Lei},
  journal={Proceedings of the ACM on Software Engineering},
  year={2026},
  publisher={ACM New York, NY, USA}
}

@misc{@unipi,
author = {UNIPI Technology},
title = {Unipi Neuron},
howpublished = {"\url{https://www.unipi.technology/?gad_source=1&gclid=EAIaIQobChMIha3lqJ6EhwMV1zYIBR3LnQXdEAAYASAAEgL9_vD_BwE}"},
note = {[Online; accessed July 2026]}
}

@inproceedings{alsabbagh2024investigating,
  title={Investigating the Security of OpenPLC: Vulnerabilities, Attacks, and Mitigation Solutions},
  author={Alsabbagh, Wael and Kim, Chaerin and Langend{\"o}rfer, Peter},
  booktitle={IEEE Access},
  year={2024},
  publisher={IEEE}
}

@inproceedings{canedo2014high,
  title={High communication throughput and low scan cycle time with multi/many-core programmable logic controllers},
  author={Canedo, Arquimedes and Ludwig, Hartmut and Al Faruque, Mohammad Abdullah},
  booktitle={IEEE Embedded Systems Letters},
  year={2014}
}

@inproceedings{florida,
  title={Dangerous stuff’: hackers tried to poison water supply of Florida Town},
  author={Robles, Frances and Perlroth, Nicole},
  booktitle={The New York Times},
  year={2021}
}

@inproceedings{@SAIN,
  title={SAIN: Improving ICS Attack Detection Sensitivity via State-Aware Invariants},
  author={Abbas, Syed Ghazanfar and Ozmen, Muslum Ozgur and Alsaheel, Abdulellah and Khan, Arslan and Celik, Z Berkay and Xu, Dongyan},
  booktitle={USENIX Security Symposium },
  year={2024}
}

@inproceedings{diedrich2004function,
  title={Function block applications in control systems based on IEC 61804},
  author={Diedrich, Christian and Russo, Francesco and Winkel, Ludwig and Blevins, Terry},
  booktitle={ISA transactions},
  year={2004}
}

@inproceedings{aoudi2018truth,
  title     = {Truth Will Out: Departure-Based Process-Level Detection of Stealthy Attacks on Control Systems},
  author    = {Aoudi, Wissam and Iturbe, Mikel and Almgren, Magnus},
  booktitle = {Proceedings of the 2018 ACM SIGSAC Conference on Computer and Communications Security},
  pages     = {817--831},
  year      = {2018},
  doi       = {10.1145/3243734.3243781}
}

@inproceedings{lin2018tabor,
  title     = {TABOR: A Graphical Model-based Approach for Anomaly Detection in Industrial Control Systems},
  author    = {Lin, Qin and Adepu, Sridha and Verwer, Sicco and Mathur, Aditya},
  booktitle = {Proceedings of the 2018 ACM Asia Conference on Computer and Communications Security},
  pages     = {525--536},
  year      = {2018},
  doi       = {10.1145/3196494.3196546}
}

@article{abdelaty2021daics,
  title   = {DAICS: A Deep Learning Solution for Anomaly Detection in Industrial Control Systems},
  author  = {Abdelaty, Maged Fathy and Doriguzzi-Corin, Roberto and Siracusa, Domenico},
  journal = {IEEE Transactions on Emerging Topics in Computing},
  volume  = {10},
  number  = {2},
  pages   = {1117--1129},
  year    = {2022},
  doi     = {10.1109/TETC.2021.3073017}
}

@inproceedings{lin2023dnattest,
  title     = {DNAttest: Digital-twin-based Non-intrusive Attestation under Transient Uncertainty},
  author    = {Lin, Wei and Tan, Heng Chuan and Chen, Binbin and Zhang, Fan},
  booktitle = {2023 53rd Annual IEEE/IFIP International Conference on Dependable Systems and Networks (DSN)},
  pages     = {376--388},
  year      = {2023},
  doi       = {10.1109/DSN58367.2023.00044}
}

@article{adis2025,
  title   = {ADIS: Detecting and Identifying Manipulated PLC Program Variables Using State-Aware Dependency Graph},
  author  = {Yang, Zeyu and He, Liang and Hu, Yujiao and Cheng, Peng and Chen, Jiming and Zhou, Jianying},
  journal = {IEEE Transactions on Information Forensics and Security},
  volume  = {20},
  pages   = {12445--12459},
  year    = {2025},
  doi     = {10.1109/TIFS.2025.3629569}
}

@inproceedings{wang2022rttee,
  title     = {RT-TEE: Real-time System Availability for Cyber-physical Systems using ARM TrustZone},
  author    = {Wang, Jinwen and Li, Ao and Li, Haoran and Lu, Chenyang and Zhang, Ning},
  booktitle = {2022 IEEE Symposium on Security and Privacy (SP)},
  pages     = {352--369},
  year      = {2022},
  doi       = {10.1109/SP46214.2022.9833604}
}

@inproceedings{gowrisankar2022gatekeeper,
  title     = {GateKeeper: Operator-centric Trusted App Management Framework on ARM TrustZone},
  author    = {Gowrisankar, Balachandar and Mashima, Daisuke and Ong, Wenshei and Ye, Quanqi and Esiner, Ertem and Chen, Binbin and Kalbarczyk, Zbigniew},
  booktitle = {2022 IEEE Conference on Communications and Network Security (CNS)},
  pages     = {100--108},
  year      = {2022},
  doi       = {10.1109/CNS56114.2022.9947233}
}
%\bibliography{reference2}
%\bibliographystyle{plainurl}
%\bibliography{\jobname}

%%%%%%%%%%%%%%%%%%%%%%%%%%%%%%%%%%%%%%%%%%%%%%%%%%%%%%%%%%%%%%%%%%%%%%%%%%%%%%%%
\end{document}